\documentclass[aps,prl,twocolumn,superscriptaddress,longbibliography,epsfig,floats,preprintnumbers,floatfix]{revtex4-1}
\usepackage{graphicx}
\usepackage[latin2]{inputenc}
\usepackage{amsmath}
\usepackage{amssymb}
\usepackage{bm}
\usepackage{color}
\usepackage{sidecap}

\newcommand{\vcst}{v_\text{c}^\text{st}}
\newcommand{\li}{\lambda_{_0}}
\newcommand{\oi}{\theta\!_{_0}}
\newcommand{\lt}{\lambda(t)}
\newcommand{\ot}{\theta(t)}
\newcommand{\bunderline}[1]{\underline{#1\mkern-4mu}\mkern4mu }

\begin{document}
\title{Geometry-Driven Segregation in Periodically Textured Microfluidic Channels}
\author{Fatemeh S. Ahmadi}
\affiliation{Department of Physics, K.N.\ Toosi University of Technology, Tehran 15875-4416, Iran}
\author{Hossein Hamzehpour}
\affiliation{Department of Physics, K.N.\ Toosi University of Technology, Tehran 15875-4416, Iran}
\affiliation{School of Quantum Physics and Matter, Institute for Research in Fundamental Sciences 
(IPM), Tehran 19538-33511, Iran}
\author{Reza Shaebani}
\affiliation{Department of Theoretical Physics and Center for Biophysics, 
Saarland University, 66123 Saarbr\"ucken, Germany}

\begin{abstract}
We investigate the transport dynamics of elongated microparticles in microchannel flows. While 
smooth-walled channels preserve the dependence of particle trajectories on initial orientation 
and lateral position, we show that introducing periodically textured walls can trigger robust 
alignment of the particle along the channel centerline. This geometry-driven alignment arises 
from repeated reorientations generated by spatially modulated shear gradients near the 
textured walls. The alignment efficiency depends on particle elongation and the relative texture 
wavelength, with an optimal range for maximal effect. While the observed alignment behavior is 
not limited to low Reynolds numbers, the characteristic alignment length scale diverges as the 
Reynolds number increases toward the turbulent flow regime. These findings offer a predictive 
framework for designing microfluidic devices that passively sort or focus anisotropic particles, 
with implications for soft matter transport, biophysical flows, and microfluidic engineering.
\end{abstract}

\maketitle

\section{Introduction}
 
Understanding particle flow in microscale environments is vital for advances in technology, 
medicine, and industry. The complexity arises from the interplay between particle properties, 
fluid behavior, and boundary conditions \cite{Squires10,Squires05,Stoecklein19,Shaebani20}. 
Particularly important is transport through narrow passages, as in blood vessels where 
lateral dispersion affects drug delivery and biomaterial transport \cite{Kainka25}. 
In physiological microcirculation, suspended cells such as red blood cells often exhibit 
spatial organization across the vessel cross section, leading to the formation of a 
cell-free layer near vessel walls and preferential focusing toward the channel centerline. 
This phenomenon has been observed in microfluidic experiments and simulations of 
blood flow in confined geometries (including constricted channels and bifurcating 
microvascular networks, where confinement and hydrodynamic interactions regulate the 
lateral distribution of cells and strongly influence transport processes) \cite{Rashidi23,
Rashidi23b,Recktenwald23}. Such studies highlight how microscale confinement and 
hydrodynamic interactions can drive particle focusing and spatial organization in 
flowing suspensions.

Broad applications in microfluidic devices span particle purification, sorting, and 
filtration \cite{Huang04,Collins14,Loutherback09,Tang22,Xu13,Sajeesh14}. Separation 
methods are broadly classified as active or passive. Active methods use external 
acoustic, electric, magnetic, or optical fields, which may pose risks to sensitive 
biomaterials, e.g.\ in cell sorting \cite{Lenshof10,Sivaramakrishnan20,Valani24,
Sailer24,acoustofluidic2023,Sajeesh14}. Passive methods leverage intrinsic interactions 
with channel architecture and flow \cite{Loutherback09,Tang22,Lenshof10,Sajeesh14,Xu13}. 
For example, topographically patterned channels can generate secondary flows such as 
microvortices, which enable size-dependent trapping and focusing by selectively 
capturing particles in recirculating regions \cite{Park09,Zhou13}.
Several studies have examined how confinement, complex channel geometries, and interfacial 
stresses influence the motion and deformation of objects in microfluidic environments, 
revealing rich transport behaviors in constrictions, junctions, and corrugated channels 
\cite{Mandal21,Mandal23,Mandal24,Mandal25}. These investigations highlight the important 
role of channel geometry, viscosity contrast, and capillary stresses in shaping transport 
processes in microfluidic systems. Related effects also arise in porous media, where flow through 
periodically constricted pathways can be modeled as transport through arrays of corrugated 
capillaries. In such systems, particle deformability, size, and concentration strongly 
influence migration, pressure drop, and effective permeability, highlighting the broader 
relevance of geometry-induced transport modulation \cite{Li21}. Among passive techniques, 
deterministic lateral displacement uses arrays of obstacles to sort spherical particles 
by size or deformability \cite{Huang04,Collins14,McGrath14,Zhbanov23,Salafi19,Li18}, 
guiding trajectories based on particle properties.

In addition to obstacle-based strategies, spatially modulated channel geometries such as 
corrugated or wavy walls have emerged as an alternative route to particle focusing. In 
such systems, periodic variations in confinement generate oscillatory acceleration and 
straining fields that can induce net cross-stream migration. Early work demonstrated 
that even for spherical particles, corrugated tubes can produce a cumulative radial 
drift leading to centerline focusing or off-axis attractors depending on Reynolds 
number and particle inertia \cite{Hewitt10}. Subsequent theoretical and computational 
studies have shown that this mechanism can be understood in terms of oscillatory 
clustering and inertial drift arising from repeated exposure to spatially varying 
flow gradients \cite{Nizkaya14,Storm24}. More recent experimental and analytical work 
has further highlighted the interplay between inertial lift forces and geometry-induced 
straining in wavy channels, revealing that the competition between these effects determines 
stable focusing locations across a wide range of flow conditions \cite{Mao23}. These studies 
establish corrugated channels as a versatile platform for passive particle manipulation based 
on flow-geometry coupling.

The dynamics of anisotropic particles in viscous flows have long been studied in the context 
of elongated or ellipsoidal particles. In unbounded shear flows, such particles follow periodic 
rotational trajectories known as Jeffery orbits \cite{Jeffery22}. In confined channels, however, 
the presence of boundaries and spatially varying shear modifies these dynamics, often leading 
to deviations from ideal Jeffery motion and generating complex translational-rotational coupling. 
Previous studies have shown that confinement and hydrodynamic interactions with channel walls 
can influence particle orientation, migration, and alignment in microfluidic environments. 
These effects are particularly relevant for elongated particles whose rotational dynamics 
couples strongly to local shear gradients, making them sensitive probes of flow structure 
and boundary-induced perturbations.

Conventional microfluidic separation devices often fail for real-world non-spherical 
particles \cite{Basagaoglu18}. I-shaped pillar arrays were used to sort disc-shaped 
soft cells \cite{Zeming13}, but no universal design exists for arbitrary shapes, and 
fabrication constraints limit applicability. Recent studies show that particles with 
a single mirror-symmetry axis can self-align to the centerline in Stokes flow \cite{Uspal13,
Georgiev20}, though boundary roughness or thermal noise can disrupt this \cite{Fiorucci19}. 
In contrast, particles with two symmetry axes \cite{Uspal13,Nagel18} or asymmetric surface 
properties \cite{Trofa19} persistently rotate and migrate laterally. Beyond smooth channels, 
microfluidic flows in structured or corrugated geometries can generate spatially modulated 
velocity fields that strongly influence particle motion. In particular, periodic wall modulations 
give rise to oscillatory shear and secondary flow structures that can drive lateral 
migration and focusing through cumulative drift mechanisms, even in the absence of 
external forcing \cite{Hewitt10,Mao23}. Such geometrical modulations have been shown 
to alter particle trajectories, enhance mixing, or induce complex transport patterns 
through periodic variations in local shear and confinement. While random surface roughness 
typically enhances dispersion and irregular wall interactions \cite{Kurzthaler20,Gamrat08,
Saccone22,AcostaCuevas23}, periodic geometric structures can instead produce deterministic 
modifications of the flow field that may systematically affect particle orientation and 
migration.

Despite these insights, most existing studies focus on spherical or weakly deformable 
particles, and the interplay between particle shape, confinement, and structured channel 
boundaries remains poorly understood. In particular, how spatially modulated shear 
fields generated by textured walls influence the orientation dynamics and transport 
of elongated particles has received little attention. This limits the development 
of efficient passive alignment and separation strategies for anisotropic microparticles.

Here, we investigate the transport of elongated particles in microfluidic channels 
with periodic boundary textures. We reveal a robust geometry-driven alignment mechanism that 
selectively guides particles toward the centerline based on their shape. Using simulations, 
we show how spatially modulated shear gradients passively drive particles toward stable, 
streamwise-oriented trajectories. This analysis yields predictive design guidelines for 
channel geometries that promote alignment for given particle shapes and flow conditions. 
Our findings offer a scalable, passive strategy for elongation-based particle focusing 
with broad relevance to soft matter, biophysics, microfluidic engineering, and materials 
science.

\begin{figure}[t]
\centering
\includegraphics[width=0.47\textwidth]{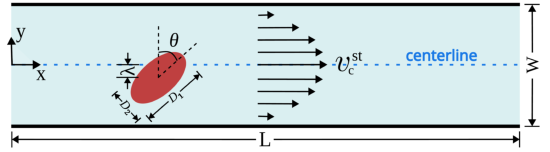}
\caption{Sketch of the simulation setup. The particle's orientation $\ot$ 
and lateral position $\lt$ generally change as the particle moves along the 
channel.}
\label{Fig1}
\end{figure}

\section{Model}

We study elongated rigid particles suspended in a steady unidirectional flow through 
a straight two-dimensional (2D) rectangular channel of length $L$ and width $W$ 
(Fig.\,\ref{Fig1}). Particles are modeled as ellipses with aspect ratio 
\begin{equation}
\kappa=D_2 / D_1,
\end{equation}
where $D_1$ and $D_2$ are the major and minor diameters. Throughout 
this study, we fix the major diameter $D_1$ and vary $D_2$ to control $\kappa$, 
allowing us to isolate the effect of particle elongation without altering the length 
scales in the system. The aspect ratio ranges from $\kappa{=}0$ (ideal rod, $D_2{=}0$) 
to $\kappa{=}1$ (disk of radius $R{=}D_1{=}D_2$). The particle's orientation $\theta$ 
is the angle between the major axis and the $y$-axis, and $\lambda$ is the lateral 
distance of its center of mass from the channel centerline.

The fluid is an incompressible Newtonian liquid with density $\rho$ and dynamic 
viscosity $\eta$. The viscous stress tensor is 
\begin{equation}
\pmb{{\bunderline{\sigma}}}={-}P\pmb{I}+2\eta\pmb{D},
\end{equation}
which is linearly related to the strain rate tensor 
\begin{equation}
\pmb{D}=\frac{1}{2}(\nabla\pmb{v}+\nabla\pmb{v}^T).
\end{equation}
Here, $P$, $\pmb{I}$, and $\pmb{v}$ are pressure, identity tensor, and fluid velocity. 
Fluid motion is governed by the Navier-Stokes and incompressibility equations 
\begin{equation}
\rho(\frac{\partial\pmb{v}}{\partial t}+\pmb{v}{\cdot}\!\nabla\pmb{v})={-}\nabla 
P+\eta\nabla^2\pmb{v},
\end{equation}
\begin{equation}
\nabla\!{\cdot}\pmb{v}=0.
\end{equation}

\begin{table}[t]
\centering
\label{Table1}
\begin{tabular}{lccc}
			\hline\hline
			Parameter&Symbol&Value&Unit\\
			\hline
			channel width & $W$ & 100 & $\mu$m\\
			channel length & $L$ & 700 & $\mu$m\\
			dynamic viscosity & $\eta$ & 1  & mPa$.$s\\
			fluid density & $\rho$ & $10^3$ &  kg/m$^3$\\				
			disk diameter & $R$ & 10 & $\mu$m\\
			ellipse major diameter & $D_1$ & 40 & $\mu$m\\			
			particle mass & $m$ & $10^{-8}$ & kg\\
			pressure difference & $\Delta P$ & 8-9 & Pa\\
			initial lateral distance & $\lambda_0$ & 0  & $\mu$m\\
			initial orientation & $\theta_0$ & 0  & rad\\
			aspect ratio (particle elongation) & $\kappa$ & 0.5  & \\	
			\hline
\end{tabular}
\caption{Set of default parameter values. Pressure difference $\Delta P$ is 
$8\,\text{Pa}$ in Figs.\,2, 3, and 5, and $9\,\text{Pa}$ in other figures.}
\end{table}
\begin{figure}[b]
\centering
\includegraphics[width=0.47\textwidth]{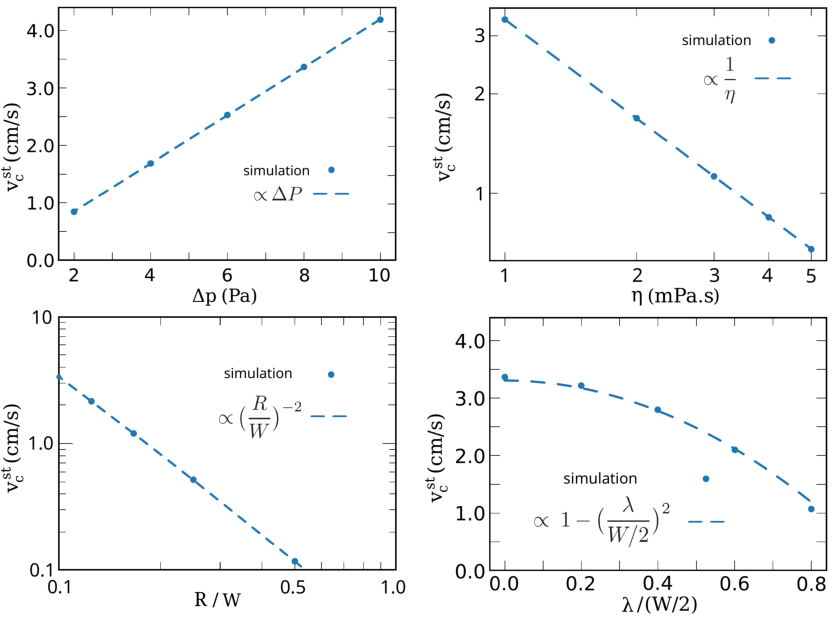}
\caption{Stationary center-of-mass velocity of a disk as a function of the 
pressure difference $\Delta P$, viscosity $\eta$, disk radius $R$, and lateral 
position $\lambda$. Default parameter values from Table I are used.}
\label{Fig2}
\end{figure}

The steady-state solution for no-slip boundary conditions at the channel walls 
($y{=}{\pm}\frac{W}{2}$) and symmetry along the centerline ($y{=}0$) is Poiseuille 
flow profile \cite{Richardson89} 
\begin{equation}
\pmb{v}^{\text{st}}(y)=\vcst\Big(1-\big(\frac{y}{W{/}2}\big)^2\Big)\pmb{\hat x},
\end{equation}
where $\vcst{=}\frac{\Delta P\,W^2}{16\eta L}$ is the maximum velocity at the centerline 
and $\Delta P$ the pressure difference between the channel inlet and outlet. The 
hydrodynamic force $\pmb{F}$ and torque $\pmb{T}$ on the particle are 
\begin{equation}
\pmb{F}=\displaystyle\int_S\;\pmb{{\bunderline{\sigma}}}{\cdot}\pmb{\hat{n}}\;ds,
\end{equation}
\begin{equation}
\pmb{T}=\displaystyle\int_S\;(\pmb{r}_s-\,\pmb{r}_{_{CM}}){\times}(\pmb{{\bunderline{\sigma}}}{\cdot}
\pmb{\hat{n}})\;ds,
\end{equation}
where $\pmb{r}_s{-}\,\pmb{r}_{_{CM}}$ connects the center of mass to a surface point 
$s$ on the particle surface $S$ with unit normal vector $\pmb{\hat{n}}$. The 
center-of-mass velocity $\pmb{v}_{_{CM}}$ and angular velocity $\omega$ are 
determined from $\pmb{F}{=}m\frac{d\pmb{v}_{_{CM}}}{dt}$ and $\pmb{T}{=}I\frac{d\omega}{dt}$ 
($I$ is the moment of inertia). Using the initial conditions $\pmb{r}_{_{CM}}(0){=}(0,\li)$, 
$\theta(0){=}\oi$ and $\pmb{v}_{_{CM}}(0){=}\pmb{0}$, we obtain the translational and 
angular dynamics of the particle 
\begin{equation}
\pmb{r}_{_{CM}}(t)=\li\,\pmb{\hat y}+\!\displaystyle\int_0^t\pmb{v}_{_{CM}}(t') dt',
\end{equation}
\begin{equation}
\theta(t)=\oi+\!\displaystyle\int_0^t\omega(t') dt'.
\end{equation}
The incompressible Navier-Stokes and continuity equations are numerically solved using 
the finite-element method (FEM) on an unstructured triangular mesh. Local mesh refinement 
is applied near the particle surface and near the textured walls, where velocity gradients 
are largest. The fluid-particle system is fully coupled:\ hydrodynamic forces and torques 
obtained from the fluid stress tensor determine the translational and rotational motion 
of the particle, while the particle motion imposes the no-slip condition on the surrounding 
fluid. This coupling is solved simultaneously at each time step to ensure momentum 
conservation and accurate fluid-structure interaction. 

To handle the moving particle, we employ an arbitrary Lagrangian-Eulerian (ALE) moving-mesh 
approach similar to standard FEM implementations. The particle boundary is treated in a 
Lagrangian frame, while the interior fluid mesh deforms smoothly in an Eulerian frame. 
Mesh displacement is propagated into the interior using a Laplace smoothing algorithm, 
with curvature-based weighting to preserve element quality near sharp features such as 
particle tips and the edges of textured walls. Mesh quality metrics, including element 
skewness and aspect ratio, are continuously monitored. If the mesh quality falls below 
a prescribed threshold, the domain is automatically remeshed, and solution fields are 
interpolated onto the new mesh. This deform-remesh-interpolate cycle ensures numerical 
stability throughout the particle trajectory, even for large translations and rotations.

\begin{figure}[b]
\centering
\includegraphics[width=0.47\textwidth]{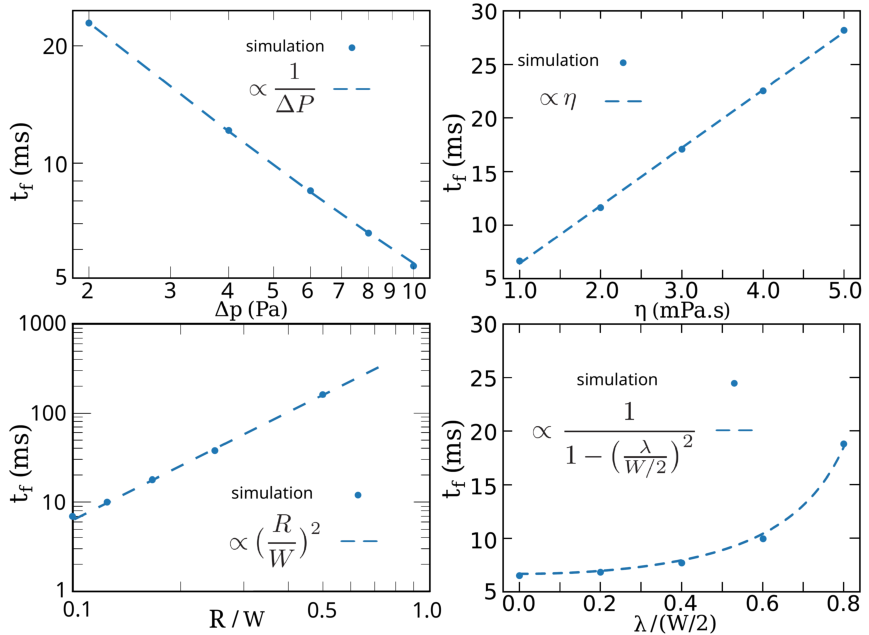}
\caption{Transit time $t\!_{_f}$ of a disk through the channel in terms of 
the pressure difference $\Delta P$, viscosity $\eta$, disk radius $R$, and 
lateral position $\lambda$. Default parameter values from Table I are used.}
\label{Fig3}
\end{figure}
\begin{figure*}
\centering
\includegraphics[width=0.9\textwidth]{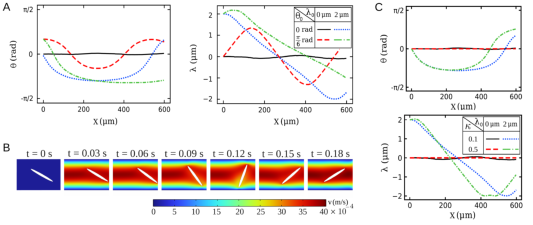}
\caption{Evolution of particle's lateral position and orientation in a smooth 
channel. (A) Orientation $\theta$ and center-of-mass distance from the centerline 
$\lambda$ as functions of the position $x$ along the channel axis for a particle 
with elongation $\kappa\,{=}\,0.1$ and $D_1\,{=}\,40\,\mu\text{m}$, $W{=}\,50\,
\mu\text{m}$, and different initial conditions $\oi$ and $\li$. (B) Snapshots 
illustrating the motion of the particle starting with $\li{=}\,0$ but $\oi{\neq}
\,0$. (C) Evolution of $\theta$ and $\lambda$ as a function of $x$ for $\oi{=}\,0$ 
and varying values of $\li$ and $\kappa$.}
\label{Fig4}
\end{figure*}

The computational domain is discretized using approximately $10^5$ triangular elements, 
with the smallest elements concentrated near the particle and wall textures. A mesh-independence 
study was conducted by varying the maximum element size and local refinement levels. The 
velocity field for a baseline Poiseuille flow was verified against the analytical solution, 
and further refinement led to negligible changes in the computed particle trajectories and 
alignment statistics. The final mesh was selected as a compromise between computational 
cost and numerical accuracy.

Time integration is performed using an implicit second-order backward differentiation formula 
with adaptive time stepping to maintain stability and accuracy, particularly when particle 
acceleration is large. At each time step, the nonlinear system arising from the FEM discretization 
is solved using a Newton-Raphson method with a direct or iterative linear solver. The overall 
approach closely follows standard FEM-ALE formulations for fluid-structure interaction problems 
as described in the literature \cite{Hughes81}. This approach provides a robust, accurate, and 
fully coupled solution of particle motion and fluid flow, while accommodating arbitrary particle 
translations and rotations within microchannels with textured boundaries. The method allows 
exploration of a wide range of Reynolds numbers and particle geometries, as presented in the 
results section.

In the low Reynolds number regime ($10^{-2}\,{\lesssim}\,R_e\,{\lesssim}\,10^{-1}$ 
in our simulations), a neutrally buoyant spherical particle released at lateral 
position $\lambda$ with zero velocity does not exhibit lateral migration. 
Its center-of-mass velocity increases over time and asymptotically approaches 
a stationary value that depends on particle size. For a point-like particle, 
the evolution follows the fluid velocity $\pmb{v}^{
\text{st}}(\lambda)$ as 
\begin{equation}
\pmb{v}(\lambda,t)=\pmb{v}^{\text{st}}(\lambda)\big(1-\text{e}^{-t{/}\tau}\big),
\end{equation}
where $\tau$ is a relaxation time. Stationary velocity $v_\text{c}^\text{st}$ and 
transit time through the channel $t\!_{_f}$ vary with particle radius $R$, $\lambda$, 
$\Delta P$, and $\eta$ (Figs.\,\ref{Fig2},\,\ref{Fig3}). Default parameter values are 
listed in Table I. An anisotropic particle, such as an ellipse, undergoes lateral 
drift even in the low Reynolds number regime due to hydrodynamic interactions induced 
by velocity gradients. We note that in the present work we solve the full 
incompressible Navier-Stokes equations in order to capture possible inertial effects 
and to assess their influence on the particle alignment mechanism.

\section{Results}

\subsection{Motion in smooth channels}

\begin{figure}[t]
\centering
\includegraphics[width=0.47\textwidth]{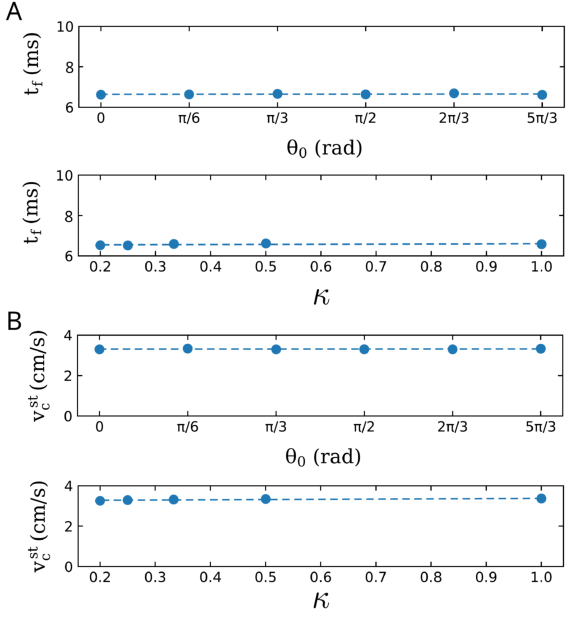}
\caption{(A) Transit time $t\!_{_f}$ and (B) stationary center-of-mass velocity 
$v_\text{c}^\text{st}$ of an ellipse as a function of the particle's initial 
orientation $\oi$ and aspect ratio $\kappa$. Default parameter values from 
Table I are used.}
\label{Fig5}
\end{figure}

In a smooth-walled channel, an elongated particle entering with symmetric initial 
conditions ($\li{=}\,0$ and $\oi{=}\,0$ or $\frac{\pi}{2}$) maintains $\lt{\approx}0$ 
and $\ot{\approx}0$ (or $\frac{\pi}{2}$) within numerical accuracy. For asymmetric initial 
conditions ($\li{\neq}\,0$, $\oi{\notin}{0,\frac{\pi}{2}}$), both $\lt$ and $\ot$ 
vary continuously, deviating from classical Jeffery orbits in unbounded flow 
\cite{Jeffery22} due to entrance-induced hydrodynamic effects (Fig.\,\ref{Fig4}A,B and 
Movie\,S1). For particles initially centered but misaligned, we observe regular 
rotational motion resembling Jeffery orbits (red dashed line in Fig.\,\ref{Fig4}A). 
Moreover, when the channel is widened and the particle is placed far from the walls, 
entrance effects diminish and trajectories converge to closed periodic orbits in 
orientation space, validating our numerics in the low-Reynolds-number regime. As 
shown in Fig.\,\ref{Fig4}C, increasing elongation $\kappa$ increases the characteristic 
length over which the particle orientation and trajectory repeat approximately, 
reflecting longer quasi-periodic cycles of motion along the channel (Exact repetition 
is limited by the finite channel length, so the particle exhibits quasi-periodic 
behavior over this characteristic length). However, $v_\text{c}^\text{st}$ and 
$t\!_{_f}$ remain largely insensitive to $\kappa$ or initial conditions; see 
Fig.\,\ref{Fig5}. The weak dependence of the center-of-mass velocity on the particle 
elongation is consistent with classical results for neutrally buoyant particles in 
creeping shear flows, where the translational motion largely follows the local fluid 
velocity at the particle's center. In contrast, the particle elongation strongly 
influences the rotational dynamics, modifying the period and spatial wavelength 
of the Jeffery-type trajectories. Within the low-Reynolds-number regime considered 
here, confinement and entrance effects do not significantly alter the translational 
speed. In the following, we show the situation changes in textured channels, where 
the particle elongation becomes a key factor governing the alignment dynamics.

\subsection{Motion in periodically textured channels}

\begin{figure}[t]
\centering
\includegraphics[width=0.47\textwidth]{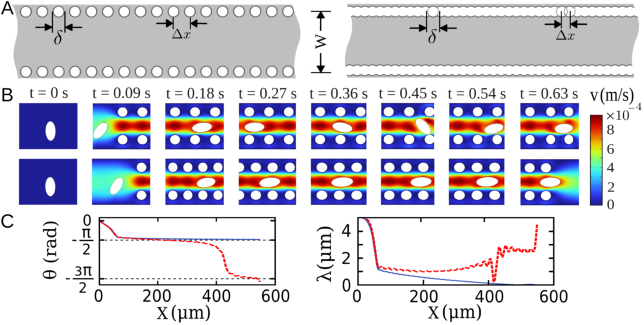}
\caption{Motion in periodically textured channels. (A) Illustrative sketches 
of periodically textured channel boundaries. (B) Time-lapse snapshots 
of an elongated particle ($\kappa\,{=}\,0.5$) moving through channels 
of width $W{/}\delta\,{=}\,3.0$. Top: For texture wavelength $\Delta x{/}
\delta\,{=}\,2.0$, the particle exhibits continuous rotation and lateral 
drift. Bottom: For $\Delta x{/}\delta\,{=}\,1.6$, the particle successfully 
aligns with the centerline and maintains a stable orientation. (C) Orientation 
$\theta$ and lateral position $\lambda$ as functions of axial position $x$, 
corresponding to the cases in (B). The solid blue curve shows successful 
alignment. The dashed red curve indicates an unsuccessful alignment case.}
\label{Fig6}
\end{figure}

Since elongated particles continuously drift and rotate in translationally symmetric 
flows, we ask whether disrupting the longitudinal uniformity can regulate their motion. 
Interestingly, we find that periodic textures along channel walls can induce particle 
alignment with the centerline within specific parameter ranges. To illustrate, we 
simulate a setup where immobile disks of diameter $\delta$ are placed along the walls 
with a texture wavelength $\Delta x$ (center-to-center distance); see Fig.\,\ref{Fig6}A. 
The particle enters the textured region from a wider, smooth-walled upstream segment. A 
pressure difference between the two ends of the region drives flow under no-slip 
conditions on the immobile disks. The simulation domain is large enough to ensure 
that entrance and exit effects at the inlet and outlet do not influence the observed 
alignment behavior. To verify this, we performed domain-size sensitivity tests by 
varying the lengths of the upstream smooth segment and the downstream outlet region 
while keeping all other parameters fixed. Beyond a buffer length of several particle 
diameters on each side of the textured region, variations in the minimum alignment 
length remained below $2\%$, and the final lateral position and orientation of the 
particle were unchanged, confirming that the alignment behavior is not affected by 
entrance or exit boundary effects. The periodic texture generates localized high-velocity 
zones along the centerline that repeatedly nudge the particle toward the centerline, 
promoting alignment downstream. Alignment success depends on the channel width $W$ 
and texture wavelength $\Delta x$. As shown in Figs.\,\ref{Fig6}B,C and Movies\,S2,\,S3, 
decreasing $\Delta x$ enhances alignment. We define successful alignment as the state 
in which the particle center of mass approaches the channel centerline ($\lt{\rightarrow}0$) 
while the particle's major axis aligns with the flow direction ($\ot{\rightarrow}{\pm}
\frac{\pi}{2}$). Quantitatively, alignment is considered achieved when the distance 
from the centerline falls below $1\,\mu\text{m}$ and the orientation deviation from 
the streamwise direction is less than $5^\circ$, both remaining within these bounds 
over a downstream distance of $100\,\mu\text{m}$.

\begin{figure}[t]
\centering
\includegraphics[width=0.47\textwidth]{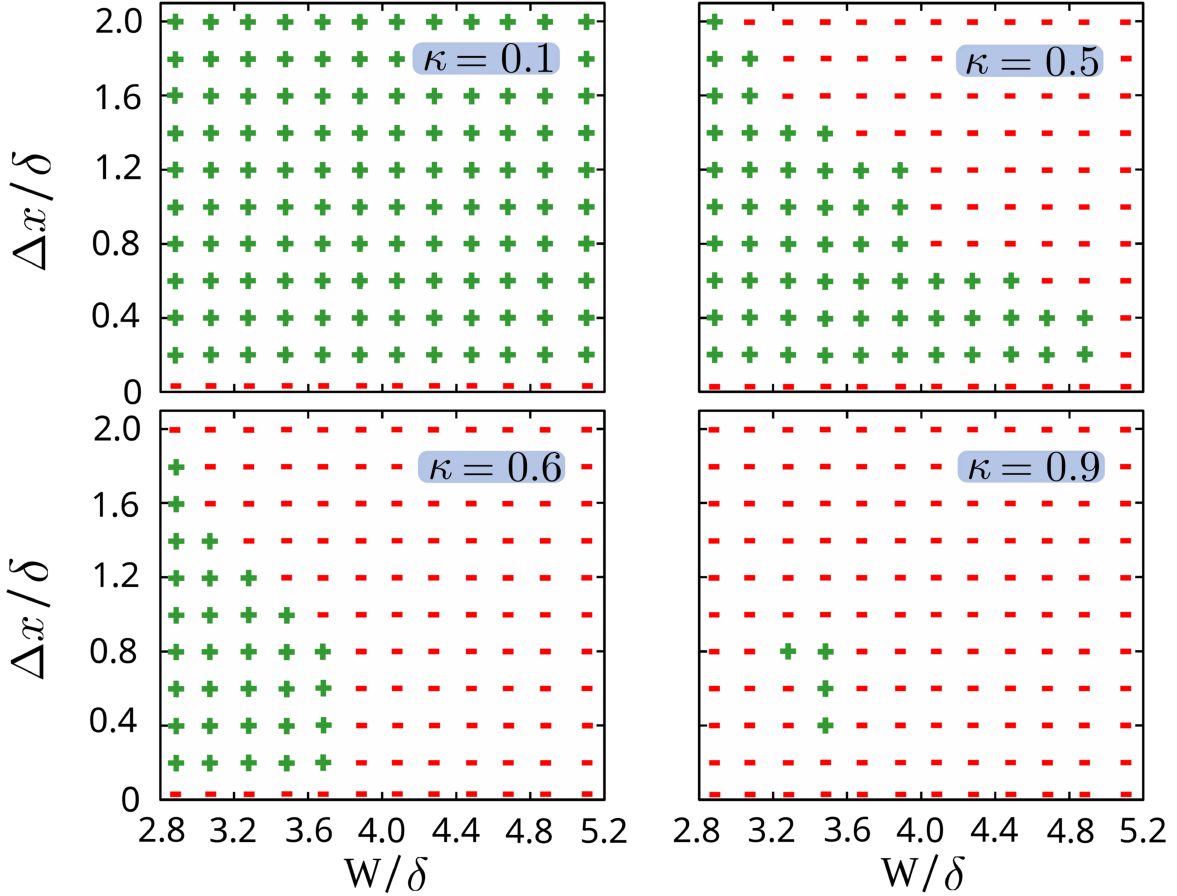}
\caption{Alignment phase diagram. Successful (green plus) and unsuccessful 
(red minus) alignment events plotted in the ($W{/}\delta$, $\Delta x{/}\delta$) 
plane for an elongated particle with a major diameter of $D_1{/}\delta\,{=}\,2$. 
The phase diagram is presented separately for different values of particle 
elongation $\kappa$.}
\label{Fig7}
\end{figure}

\begin{figure}[t]
\centering
\includegraphics[width=0.47\textwidth]{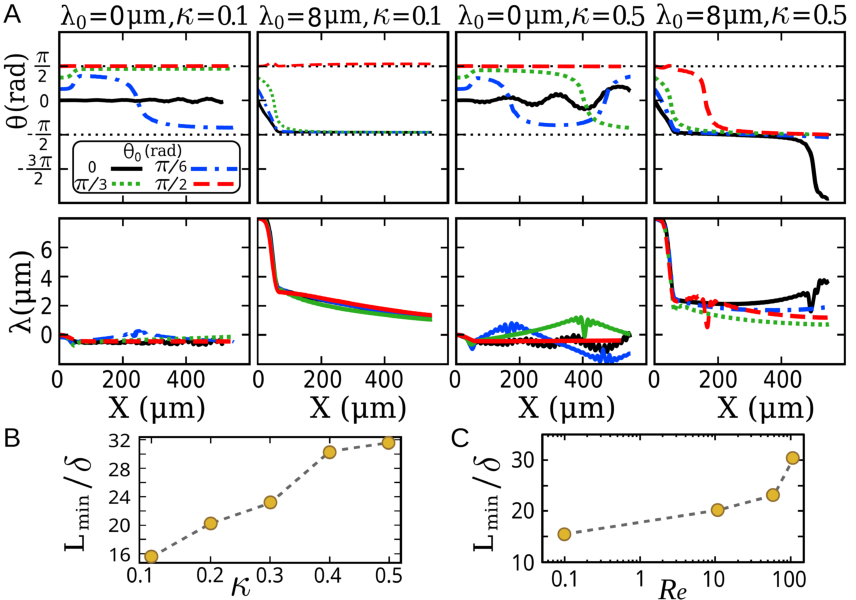}
\caption{Evolution of particle's lateral position and orientation 
in periodically textured channels. (A) $\theta$ and $\lambda$ versus 
axial position $x$ for $W{/}\delta{=}4.0$, $\Delta x{/}\delta{=}1.5$, 
and different values of $\kappa$, $\li$, and $\oi$. (B) Minimum channel 
length required for alignment, $L_\text{min}$, vs $\kappa$ for the same 
channel geometry as in (A), with $\li{=}0$ and averaged over $\oi$. (C) 
Dependence of $L_\text{min}$ on the Reynolds number for $\kappa{=}0.1$, 
$\oi{=}\frac{\pi}{3}$, and $\li{=}0$.}
\label{Fig8}
\end{figure}

Our key finding is that particle elongation enhances alignment success. 
The alignment phase diagrams in Fig.\,\ref{Fig7} reveal 
that increasing elongation expands the parameter space for 
successful alignment. This is because a more elongated particle samples 
the spatial shear gradients more effectively, thereby experiencing stronger 
reorientation torques (see Fig.\,\ref{Fig8}A). Similar high-velocity zones 
can also be created using alternative techniques such as acoustic waves 
\cite{Jannesar21,Eslami23}. Notably, alignment depends nonmonotonically 
on the texture wavelength $\Delta x$. When $\Delta x{\ll}D_1$, torques 
average out along the particle length, suppressing alignment. For $\Delta x
{\gg}D_1$, the spacing between shear zones is too large for the particle 
to experience continuous reorientation, again weakening alignment. The 
optimal regime lies near $\Delta x\,{\sim}\,D_1$, specifically $0.5 D_1
{\lesssim}\Delta x{\lesssim}2 D_1$. Alignment also depends on boundary 
roughness, captured by the dimensionless ratio $\varepsilon_1{=}\frac{\Delta x}{\delta}$. 
For $\varepsilon_1{\ll}1$, strongly overlapping disks smooth the wall and reduce 
shear gradients; for $\varepsilon_1{\gg}1$, shear zones become widely separated 
and alignment degrades. Optimal alignment occurs for $0.1{\lesssim}\varepsilon_1{\lesssim}2$, 
with effects vanishing beyond $\varepsilon_1{>}5$. Another key parameter is 
relative channel width, defined as $\varepsilon_2{=}\frac{W}{\delta}$. Strong confinement 
($\varepsilon_2{\leq}1$) limits structured shear and lateral migration and suppresses 
shear-driven alignment; for weak confinement ($\varepsilon_2{\gg}1$), the 
particle can move far from both walls, reducing exposure to the boundary-induced 
shear gradients that drive centering. Alignment is most effective for $2{\leq}
\varepsilon_2{\leq}5$. These results offer practical design guidelines for tailoring 
microchannel textures to specific particle geometries. In addition, the particle size 
relative to the channel width provides an important constraint for the alignment mechanism. 
Efficient alignment requires the particle to experience spatial variations of the shear 
field along its length; if the particle is much smaller than the channel width, the 
velocity field across its body becomes nearly uniform and the resulting reorientation 
torques weaken substantially. Consistent with our numerical observations, alignment 
disappears for $D_1{/}W\,{\lesssim}\,0.01$, while robust alignment is typically 
observed only for substantially larger particles, approximately $D_1{/}W\,{\gtrsim}
\,0.2$. In the limit of very small particles, the dynamics approaches that of 
point-like tracers that simply follow streamlines.

Particle elongation also affects the minimum travel distance $L_\text{min}$ 
required for alignment. As shown in Fig.\,\ref{Fig8}B, $L_\text{min}$, 
averaged over initial conditions, roughly doubles as $\kappa$ increases 
from $0.1$ to $0.5$. To explore finite Reynolds number 
effects on alignment behavior, we vary the dynamic viscosity and fluid 
velocity, and solve the full incompressible Navier-Stokes equations.
Figure\,\ref{Fig8}C shows that alignment is not limited to highly viscous 
flows and persists beyond the low-Reynolds-number regime: for $\kappa\,{=}\,0.1$, it 
remains effective up to $R_{e}$ on the order of a few hundred (which 
approaches turbulent flow), though $L_\text{min}$ increases. For 
larger $\kappa$, alignment remains robust for $R_{e}\,{\sim}\,1{-}10$, 
the typical range in microfluidic devices.

Importantly, the alignment in textured channels is not merely due to flow 
focusing, as would occur in a narrower smooth channel. Instead, it 
results from periodic spatial variations in the flow that induce 
repeated orientation corrections, progressively steering elongated 
particles toward centerline alignment. This mechanism differs 
fundamentally from random surface roughness effects \cite{Kurzthaler20,
Gamrat08,Saccone22,AcostaCuevas23}, which enhances dispersion and 
complex wall interactions but does not produce systematic alignment.

\subsection{Elongation-induced segregation and applications in microparticle filtering} 

\begin{figure}[t]
\centering
\includegraphics[width=0.47\textwidth]{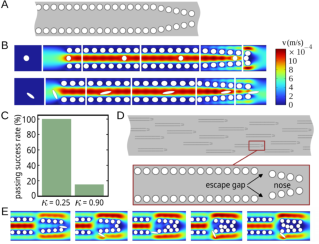}
\caption{Elongation-induced segregation and filtering. (A) Schematic of a 
textured microchannel featuring a narrow outlet bottleneck. (B) Particle 
trajectories for $\kappa{=}0.9$ (top) and $\kappa{=}0.25$ (bottom). (C) 
Passage success rates from 64 initial conditions, comparing the two elongations 
in panel (B). (D) Concept of a multi-unit filtering device, each equipped 
with an escape gap and a front nose designed for particle trapping and storage. 
(E) Time-lapse images showing elongated particles exiting while rounder ones 
accumulate at the nose.}
\label{Fig9}
\end{figure}

The dependence of the minimum alignment length $L_\text{min}$ on $\kappa$ 
enables geometry-based microparticle separation. To demonstrate elongation-induced 
segregation, we modify the textured channel by adding a narrow outlet bottleneck 
(Fig.\,\ref{Fig9}A) with width equal to the immobile disk diameter $\delta$. 
Only particles with minor diameter $D_2{<}\delta$ can pass through, if they 
align with the centerline. As shown in Fig.\,\ref{Fig9}B (and Movies\,S4,\,S5), 
highly elongated particles align quickly and pass reliably, while nearly round 
ones reach the bottleneck misaligned and are blocked. For example, particles 
with $\kappa{=}0.25$ always pass (given sufficient channel length and proper 
texture), but the success rate drops to ${\sim}15\%$ for $\kappa{=}0.9$ 
(Fig.\,\ref{Fig9}C).

To prevent clogging by large particles, we introduce a nose-shaped design with a 
narrow central outlet and a lateral escape gap (Fig.\,\ref{Fig9}D).This allows 
elongated particles to exit while trapping more-rounded ones at the nose with controlled 
storage capacity. This setup enables passive, shape-selective filtering: elongated particles, 
aligned by the upstream periodic texture, follow streamlines that bend into the side gap 
at the onset of the nose, allowing them to exit efficiently. In contrast, more-rounded 
particles exhibit broader lateral wandering and random orientations, making them less 
likely to reach the gap and more likely to be trapped at the tip or exit through the 
main outlet if it is wide enough. Thus, the filter relies on geometry-dependent flow 
alignment and redirection. Figure\,\ref{Fig9}E and Movie\,S6 illustrate this 
behavior, with elongated particles exiting while round ones accumulate. The design 
serves as a basic unit of a larger filtering device, in which a mixture of particles 
with varying aspect ratios is processed such that predominantly elongated particles 
reach the outlet. The device efficiency depends on the storage capacity of each unit 
and the aspect ratio contrast among the input particle mixture.

Beyond this conceptual demonstration, the proposed mechanism provides a general design 
principle for passive microfluidic separation devices based on particle shape. By tailoring 
the periodic texture and outlet geometry, the alignment dynamics can be tuned to selectively 
guide particles with desired aspect ratios toward specific outlets or storage regions. Such 
geometry-driven control of particle trajectories may offer a scalable strategy for shape-selective 
filtration and sorting in biomedical diagnostics, materials processing, and microfluidic 
sample preparation.

Although our simulations are 2D, this does not substantially limit the 
applicability of the results. Many microfluidic devices employ shallow, nearly 
rectangular channels fabricated by techniques such as soft lithography or PDMS 
molding, where the flow in the central region is well approximated by 2D models. 
Moreover, the underlying alignment mechanism is not inherently restricted to 2D 
and may be extended to 3D geometries using appropriately structured wall features, 
though the effects of fully 3D channel features on shear gradients and particle 
alignment warrant detailed investigation.

Before concluding, we briefly comment on the physical consistency and broader context 
of our results. The numerical framework employed here reproduces well-established 
limiting cases of particle dynamics in viscous shear flows, including the classical 
Jeffery orbits observed in smooth channels away from boundaries. Moreover, the 
simulations exhibit physically consistent behavior in several asymptotic limits 
of the model parameters, such as the recovery of the smooth-channel dynamics as 
wall roughness vanishes and the disappearance of alignment when the texture 
wavelength or channel width becomes very large compared with the particle size. 
At the same time, experimental studies of red blood cell transport in microfluidic 
channels and microvascular networks have reported strong particle focusing and the 
formation of cell-free layers near channel walls, demonstrating that confinement-driven 
particle organization is a robust phenomenon in real flows. In this context, the 
geometry-driven alignment mechanism identified here provides a predictive framework 
that may help guide future microfluidic experiments aimed at controlling the transport 
and orientation of anisotropic particles.

\section{Conclusion}

In summary, we have demonstrated that periodic texturing of microfluidic channel 
boundaries can induce alignment of elongated particles with the channel centerline. 
This effect strongly depends on particle elongation, with more elongated 
particles aligning more efficiently over shorter distances. The phenomenon 
persists across a range of Reynolds numbers, extending beyond the low-Reynolds-number 
regime, with diminishing effects at high $R_{e}$ towards turbulent flow regime due to 
inertial drift and loss of streamline coherence. These findings have significant 
implications for microfluidic applications, particularly in passive particle 
sorting and filtering technologies. Given advances in microfabrication, our 
approach is experimentally feasible, as micron-scale textures can be routinely 
fabricated using, e.g., photolithography or soft lithography. The present study 
considers dilute suspensions of monodisperse rigid particles in a Newtonian fluid. 
In practical systems, additional factors such as particle-particle interactions 
at higher concentrations, polydispersity in size or aspect ratio, particle 
deformability, and non-Newtonian rheology of the carrier fluid may influence 
the alignment dynamics. While these effects may modify alignment efficiency 
or timescales, the underlying mechanism of shear-gradient-induced reorientation 
is expected to remain relevant. Exploring these factors, together with optimizing 
texture geometries and investigating inertial effects at higher Reynolds numbers 
(e.g., inertial lateral focusing \cite{Segre61}), represents an important direction 
for future work and could further enhance the applicability of this approach 
in biomedical and industrial settings. 

\bibliography{Refs-Microfluidic}

\end{document}